\documentclass{aastex}
\usepackage{spr-astr-addons}
\usepackage{url}
\usepackage{amssymb}
\usepackage{amsthm}
\def\gas{\mathrel{\hbox{\rlap{\hbox{\lower3pt\hbox{$\sim$}}}\hbox{\raise2pt\hbox{$>$}}}}}
\def\las{\mathrel{\hbox{\rlap{\hbox{\lower3pt\hbox{$\sim$}}}\hbox{\raise2pt\hbox{$<$}}}}}

\RequirePackage{color}

\begin{document}

\title{Influence of the Centaurs and TNOs on the main belt and its families}
\slugcomment{Not to appear in Nonlearned J., 45.}
\shorttitle{Scattering of main belt asteroids by Centaurs and TNOs}
\shortauthors{Galiazzo et al.}

\author{Mattia A. Galiazzo\altaffilmark{1,2,3}} \and \author{Paul Wiegert\altaffilmark{1,2}}
\affil{Department of Physics and Astronomy, The University of Western Ontario, London, Ontario, N6A 3K7, Canada}
\and
\author{Safwan Aljbaae\altaffilmark{4}}
\affil{UNESP, Univ. Estadual Paulista, Grupo de din\^{a}mica Orbital e
  Planetologia, Guaratinguet\'{a}, SP, 12516-410, Brazil}

\altaffiltext{1}{Department of Physics and Astronomy, The University of Western Ontario, London, Ontario, N6A 3K7, Canada.}
\altaffiltext{2}{Institute of Astrophysics, University of Vienna, Turkenschanzstr. 17, A-1180 Vienna, Austria.}
\altaffiltext{3}{Centre for Planetary Science and Exploration (CPSX), London, Ontario, Canada, N6A 3K7.}
\altaffiltext{4}{TUNESP, Univ. Estadual Paulista, Grupo de din\^{a}mica Orbital e
  Planetologia, Guaratinguet\'{a}, SP, 12516-410, Brazil.}

\begin{abstract}
Centaurs are objects whose orbits are found between those of the giant planets.
They are supposed to originate mainly from the Trans-Neptunian objects, 
and they are among the sources of Near-Earth Objects.   
Trans-Neptunian Objects (TNOs) cross Neptune's orbit and produce the Centaurs. 
 We investigate their interactions with main belt asteroids to determine
 if chaotic scattering
 caused by close encounters and impacts by these bodies may have played
 a role in the dynamical evolution of the main belt. 
We find that Centaurs and TNOs that reach the inner Solar System can modify
the orbits of main belt asteroids, though only if their mass is of the
order of $10^{-9} m_\odot$ for single encounters or, one order less in
case of multiple close encounters. Centaurs and TNOs are unlikely
 to have significantly dispersed young asteroid families in the main belt,
 but they can have perturbed some old asteroid families.
Current main belt asteroids that originated as Centaurs or
 Trans-Neptunian Objects may lie in
the outer belt with short lifetime $\leq 4 My$,
  most likely between 2.8 au and 3.2 au  at
larger eccentricities than typical of main belt asteroids.
\end{abstract}



\section{Introduction}\label{sec: intro}

Centaurs are objects whose orbits are contained between those of Jupiter and 
Neptune (Gladman et al. 2008). Currently\footnote{June 2015}, 301 Centaurs
are known (JPL Small-Body Database Search Engine, JPL-SBDSE, at
http:$//$ssd.jpl.nasa.gov/sbdb\_query.cgi) and recent observational results
 estimate a value of $240\pm130$ with an absolute magnitude
 $H<10$ (Adams et al. 2014). The Centaurs' population  with a diameter
 larger than 1~km is estimated to be between $\sim10$ million (Volk and Malhotra 2013)
 and about  $\sim 8 \times 10^9$ (Fern\'andez et al. 2004, di Sisto and Brunini 2007, Napier 2015).
 No well-characterized survey has been carried out to date that
 would allow derivation of a de-biased centaur size distribution
 (Fuentes et al. 2014). 
They are thought to originate mainly from the Trans-Neptunian objects (Levison and Duncan 1993, 1997; Tiscareno and Malhotra 2003; Lykawka et al. 2009; di Sisto et al. 2010; Brasser et al. 2012) and they are among
 the sources of Near-Earth Objects (Morbidelli 1997; Levison and Duncan 1997; Tiscareno and Malhotra 2003; di Sisto and Brunini 2007).
Since some Centaurs and their progenitors can be relatively massive
 \footnote{The majority of the observed Centaurs and TNOs,
 have smaller masses, e.g. for Schwamb et al. (2014),
 large bodies comparable to Pluto sizes (and so masses) are $\sim 12$
 but in the past much more (Brasser and Morbidelli 2013).}
 ($m \gas 10^{-9} m_\odot$, of the order of one-tenth of the lunar mass,
 $\sim 4\cdot 10^{-8}$ $m_\odot$), move 
throughout the planetary system, and have done so throughout its 
existence (also in much larger numbers in the past), we ponder 
whether close encounters and  possible 
impacts by them on
 main belt asteroids may have played a role in the 
recent (that is, after the Late Heavy Bombardment, from 3.8 Gyrs ago  to now)
 dynamical evolution of the main belt and
 particularly asteroid families. 

In order to do this we consider two time spans. The first, 
which we call present population (PP),  reaching 50 Myr into the past,
 examines the effect of Centaur encounters on young asteroid families,
 i.e. the Karin family  (5.3 Myr,  Nesvorn\'y et al. 2002).
 The second, which we will call the ancient population, AP,
 stretches back 3.8 Gyrs ago  (Hartmann et al. 2000; Kirchoff et al. 2013), 
the estimated age of the end of the Late Heavy Bombardment ($LHB$) process,
 and examines the effects on  old asteroid families,
 e.g. Flora family, which is old $\sim4.4$ Gyr (Carruba et al. 2016). 
  In particular, we plan to investigate if close encounters
 with Centaurs and TNOs (from now on C+TNOs),
 with a diameter larger than 100 km\footnote{The diameter was computed using the equation of Tedesco et al. (1992), assuming the average albedo $\rho_V=0.05$ of
the Centaurs (Rabinowitz et al. 2007): $D =\frac{1329}{\rho_V}10^{-\frac{H}{5}}$. In this way, we can assume a body is larger than
100 km when its absolute magnitude, $H\las9$.},
 could have been responsible for perturbing or diffusing
 young and/or old main belt families (Zappal\'a et al. 1995; Migliorini et al. 1995; Nesvorn\'y 2012; Novakovi\'c et al. 2011), an example is
 the scattering of V-type (basaltic) asteroids from the Vesta family
 beyond the 3J:1 mean motion resonance, into
the central and outer main belt, see also Carruba et al. (2014) and Huaman et al. (2014).
 We also consider whether Centaurs contribute to the presence of interlopers inside families, like 
 the case of the C-types (carbonaceous asteroids) in the Hungaria family (up to 6\%, Warner et al. 20099), whose member are in majority E-types.\\

Centaur orbits are dynamically unstable, with a dynamical lifetime from 
less than one to about 100 Myr (Horner et al. 2004a; Tiscareno and Malhotra 2003; Bailey and Malhotra 2009).  In fact,
their dynamical evolution is mostly influenced by  close encounters with 
giant planets (as found in Horner et al. 2004a, 2004b).  
 Their lifetimes are short in general compared to
the lifetime of main belt asteroids because Centaurs are mostly 
in very chaotic regions perturbed by the giant planets. 
According to Bailey and Malhotra (2009), the lifetime of the Centaurs (at least 
the most chaotic ones, the majority) is about 22 Myrs; Dones et al. (1996) assert $\sim5$ Myr; and up to 72 Myr for di Sisto and Brunini (2007).
 Hence, they are typically removed from the Solar System on timescales of
 only millions of  years (Levison and Duncan 1997; Tiscareno and Malhotra 2003; Horner et al. 2004a and previously
 cited papers). Their lifetime typically ends by ejection on a hyperbolic
 orbit or collision with a planet (Horner et al. 2004a, 2004b; di Sisto et al. 2010; Galiazzo 2013).
  Some may become short-period  comets and a fraction of them may impact
 the terrestrial planets or the Sun.\\  

The Centaur progenitors, the TNOs, are mainly subdivided into  4
 principal regions (Adams et al. 2014): (1) \emph{Resonant} objects,
 bodies which occupy a mean-motion resonance with Neptune, e.g. the Plutinos
 (objects in N2:3 resonance with Neptune). (2) \emph{Scattered Near} objects,
 which have a semi-major axis that vary of
 $\delta a=\frac{a_{max}-a_{min}}{<a>}\geq 0.02$, over 10 Myr time scales.
 (3) \emph{Scattered Extended} objects, bodies with
 $s=\sqrt{e^2+sin^2(i)}\geq0.25$, where $e$ and $i$ are respectively
 the eccentricity and the inclination. (4) \emph{Classical} Objects,
 which have $s<0.25$.
All the dynamically excited objects with a semi-major axis,
 $a_{Nep}\las a\las 80$ au are grouped together in one single
 \emph{Scattered} class. 
  In this work we consider only the TNOs with a semi-major axis,
 $a\las 80$ au, a relatively well-characterized source region
(Adams et al. 2014) inside the heliopause. 

This paper is organized as follow: the model is described in Section 2. 
Section 3 is subdivided into 2 subsections: the semi-analytical and analytical 
studies in subsection 3.1, and in subsection 3.2 we revisit
 the numerical results. Finally, our conclusions are given in Section 4.

\section{Model}

\label{model}

To investigate how these large bodies might perturb asteroids
 in the main belt (MBAs), we performed 
 a numerical simulation of the orbital evolution of the C+TNOs
 through the main belt using the Lie-integrator  (Hanslmeier and Dvorak 1984). This 
 orbital numerical integrator has an adaptive
 step-size (Eggl and Dvorak 2010; Bancelin et al. 2012) and it was
 used in several previous works
 dealing with close encounters of asteroids with the planets
 (Galiazzo 2013; Galiazzo et al. 2013a, 2013b, 2014 and Galiazzo and Schwarz 2014).
   In this work
 we use an accuracy parameter set to $10^{-13}$.
 The output stepsize for the numerical integration is set to 1 kyr
 and a simplified solar system (SSS: all planets from Venus
 to Neptune, with the mass of Mercury added to the mass of the Sun) is
  considered for the orbital propagation.
We assume a close encounter ($CE$) is occurring when a body is within
 $\Delta r_{CE}=0.0025$ au from the perturbing  body.  
Elements of the two bodies are registered and the perturbation is 
computed from the moment the asteroid enters in this sphere of
 radius $\Delta r_{CE}$ around the massive body until the moment that it exits.
 We chose a $\Delta r_{CE}$ value roughly in the proximity of 
 the Hill's sphere radius, considering the largest Centaur:
 Centaurs Hill spheres can reach
 $R_H\sim a(1-e)(m_c/m_\odot)^{1/3}\sim 2\cdot 10^{-3}$ au, which is the
Hill sphere for the largest Centaur 1995 SN$_{55}$.


The orbital evolution of a body is considered until it either escapes
 or it collides. We consider an escape when the body has an
instantaneous eccentricity $e>0.99$, its period exceeds 1000 yrs and 
 $a\gas80 au$.  Then in order to consider the influence of a massive C+TNO 
 in the belt, we check its influence
 once its perihelion drops below 3.8 au
and if it remains there for at least 2 kyrs (2 output times steps
 of our simulations). 
For each C+TNO that meets this criterion in our initial
simulations, we make a second integration with a finer output interval
for the entire period that the body was crossing the asteroid belt, using
the elements of the 100-km sized C+TNO just as it enters the
main belt  from the first integration.

This finer integration proceeds for the duration of the C+TNO's 
residence in the main belt, as long as its perihelion is in the range 
$1.78 au < q < 3.8 au$. This outer border is chosen near the aphelion
 of 522 Helga\footnote{In practice, we examined all those below a value of 3.91 au (Helga's aphelion), to
be sure that the Centaurs and TNOs interact with this region}, the main body of the outermost asteroid family (Carruba et al. 2015) 
 and the inner one is the beginning of the Hungaria family
 (Galiazzo et al. 2013a), the closest main belt family to Mars.
 During the finer integration, all close approaches to asteroids
 are examined in order to determine the perturbation to the asteroid's orbit.

A main belt family can potentially lose one of its members  once
 a close encounter causes the asteroid (member of the family) a real
 semi-major axis change of more than  0.0003 au (from now on $K_{kill}$),
 because this is enough to affect the proper semi-major axis
at a level that may affect the apparent membership of the asteroid in 
an asteroid family (Knezevi\'c and Milani 2003). 
We will assume that the main belt (and so the asteroid belt families)
 is significantly perturbed by the $C+TNOs$ if at least $\sim 3-5\%$
 of the members of the main belt are deflected by an amount
 $\Delta a\geq K_{kill}$.

The effects of the C+TNOs on the asteroid belt are determined from their
 perturbations on a sample asteroid population.
 We want to check in particular if the TNOs are capable of scattering
  members of the Vesta Family,
 and therefore explain
 the presence of some V-types in other families
 of the belt. So we divide the asteroid belt into 2 sub-samples:
 \begin{enumerate}
\item a complete known Vesta family (Vestoid elements were taken from
 the DPS catalogue Nesvorn\'y 2012: 528 Vestoids);
\item a sample of the remaining (non-vestoid) main belt asteroids:
only a sub-sample of the remaining asteroids was propagated in order
 to save computational time.
This sample has 1054 bodies, proportionally distributed among the
  respective different regions. For our purposes,
 the main belt is divided into 6 regions, based on semimajor axis
(Table ~\ref{Tab: mbaregions}) and inclination ($i=17.16^\circ$, this value
 distinguishes the high inclined asteroid families from the low ones, as suggested in 
 Novakovi\'c et al. 2011) similarly to Galiazzo et al. (2013b).

\begin{table}[!h]
\begin{center}
\caption{Subdivision of the Main belt in different regions in ($a-i$) space
(semi-major axis interval ($\Delta a$) and inclination ($i$)).
 The number of asteroids
 (considered in the orbital integrations) which 
 belong to each region is given: ``\# low'' (low inclinations)
 stands for the number
 of asteroid in the relative semi-major axis subregion
 with inclinations $i\leq17.16^\circ$ and
 ``\# high'' (high inclinations) vice-versa,  $i>17.16^\circ$.
 The most populated region is the central belt.
MB stands for main belt, $I$, $M$ and $O$ stand respectively for
 Inner, Middle and Outer.}
\begin{tabular}[h]{|l|c|c|c|}
\hline
\hline
GROUP &   $\Delta a$  [au]   & \# low  & \# high \\
\hline
\hline
IMB &  $1.78\leq a \leq2.5$ &146 & 29\\
\hline
MMB&   $2.5< a \leq2.83$ &306 & 19\\
\hline
OMB   &   $2.83< a \leq3.8$&476 &72 \\
\hline
\hline
\end{tabular}
\label{Tab: mbaregions}
\end{center}
\end{table}

\end{enumerate}
For the size distribution of the main belt asteroids, we considered only
 bodies with diameter as large or larger than 1 km
 (with  $H<14$, asteroids with $D\approx4-9$ km,
 see http://www.minorplanetcenter.net/iau/Sizes.html)
 in order to have a representative sample quantitatively
 large enough to make a good statistical study,
 but that would not cost too much in terms of computational effort. 
 Asteroids in this size range are large enough that we can propagate
 their orbits in the belt with only gravitational forces for the small
 amount of time (usually $t < 0.5$ Myrs) a C+TNO typically interacts
 with the main belt.
Osculating elements are taken from the JPL-SBDSE which give in total
  26711 MBAs  asteroids with  $H<14$ and, as we said
before, from them we take a reasonable proportional distributed sub-sample
 for the non-vestoids and all the vestoids in this size-range.

\subsection{Present and past population} 
As mentioned earlier, we consider 2 
 populations in order to study
 the effect on the present solar system
 and on the past population just after the LHB (post-LHB).

\begin{figure}[!h]

  \centering
  \centering \includegraphics [width=0.45\textwidth]{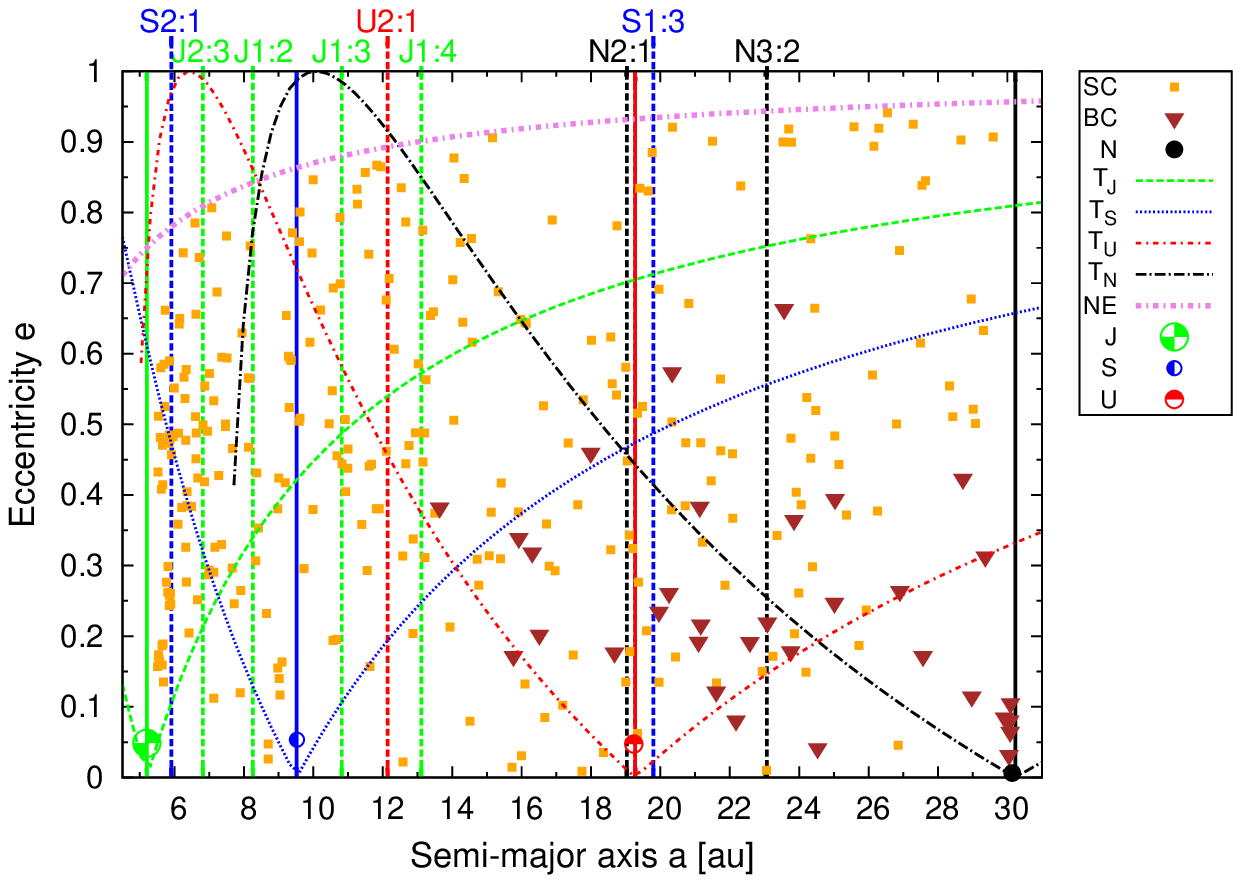}
\includegraphics [width=0.45\textwidth]{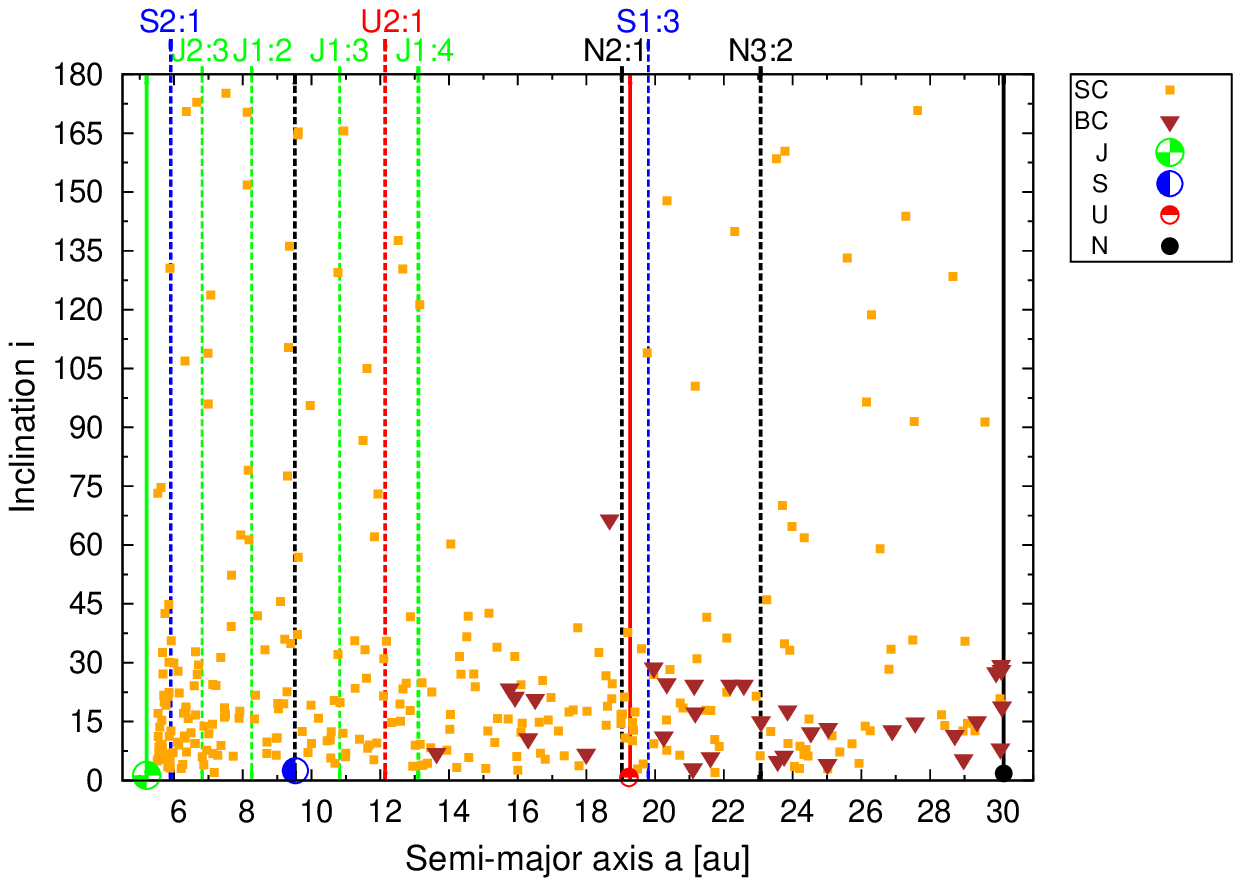}
\caption{Top panel: a semi-major axis versus eccentricity projection
 of ``small'' Centaurs (SC) and $D>100~km$ Centaurs
 (BC). 
 Vertical lines are the principal mean motion resonances (continuous
 lines are only for 1:1) for Centaurs
 with respect to the giant planet, represented by the filled circles whose
 symbols have sizes proportional to the real diameter of the planets
 (Jupiter (J), Saturn (S), Uranus (U) and Neptune (N)).
 The top curve (NE) is the border between the NEAs and the Centaurs.
 Other curved lines are the regions strongly dominated by the gravitational
 influence of the giant planets,
 represented by the Tisserand parameter curves ($T_J$, $T_S$, $T_U$ and $T_N$,
 respectively for Jupiter, Saturn, Uranus and Neptune).
Bottom panel: a semi-major axis versus inclination projection of Centaurs.
}
\label{fig: ae}
\end{figure}

\begin{figure}[!h]

  \centering
  \centering \includegraphics [width=0.45\textwidth]{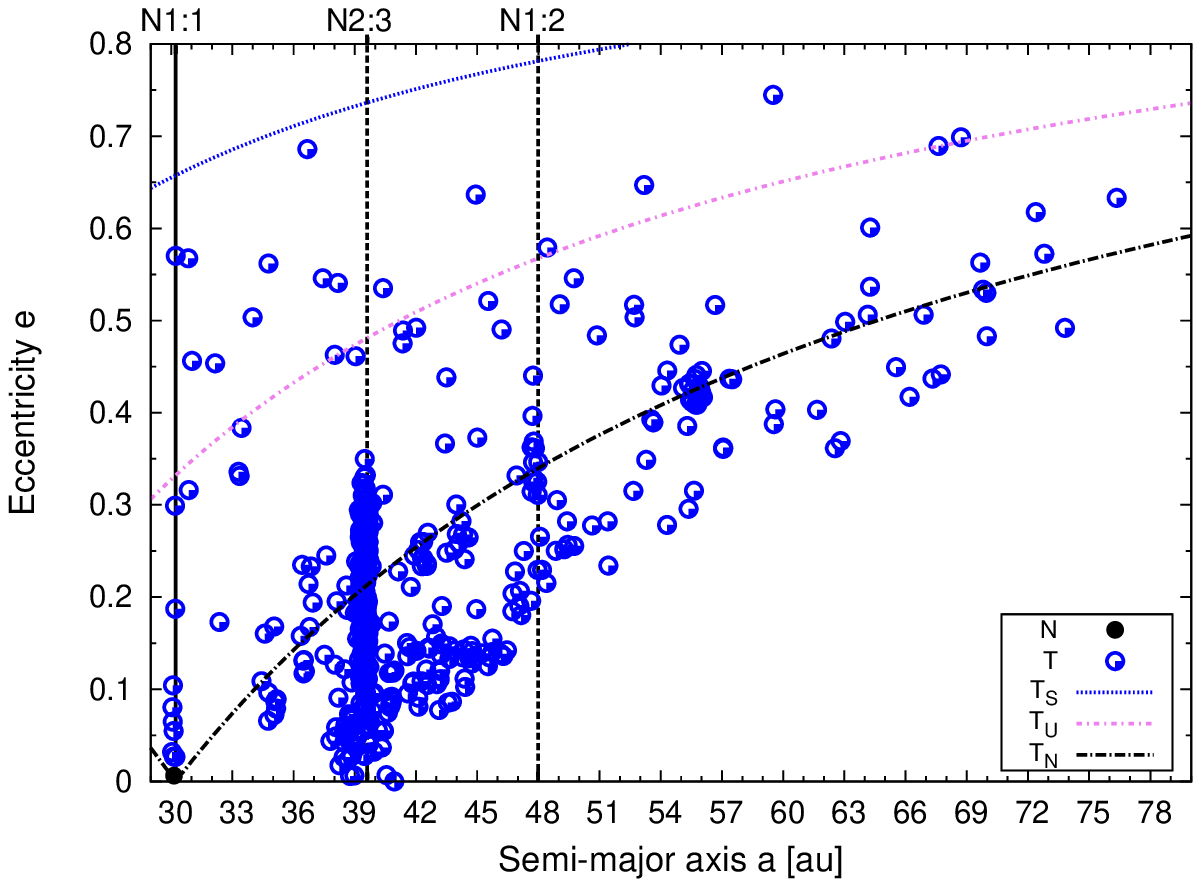}
\includegraphics [width=0.45\textwidth]{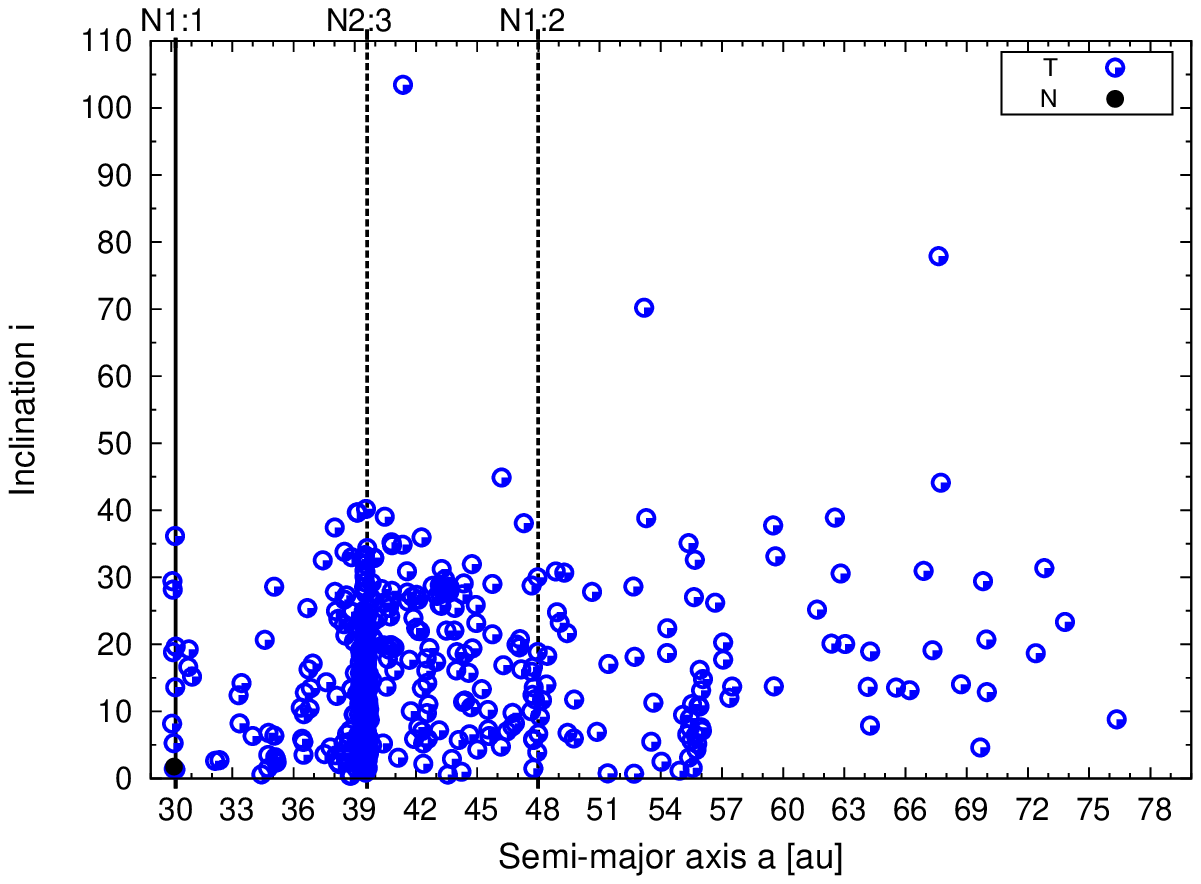}
\caption{Top panel: A semi-major axis versus eccentricity projection of TNOs
 with $D>100~km$ (T). Vertical lines represent the principal 
mean motion resonances with Neptune (N). $T_S$, $T_U$ and $T_N$ like Fig. 1.
Bottom panel: A semi-major axis versus inclination projection of TNOs
($D>100~km$) orbits.}
\label{fig: cenae}
\end{figure}

 For the $PP$
 we take as a sample the observed C+TNOs
larger than 100 km in diameter\footnote{Computed via
 the Tedesco et al. (1992) equation,
 assuming the average albedo $\rho_V=0.05$  of the Centaurs (Rabinowitz et al. 2007):
$D=\frac{1329}{\rho_V}10^{-\frac{H}{5}}$. $H$ is the absolute magnitude,
 and $D\gas 100$ when $H\las9$.} (shown in the
$a-−e$ and $a-−i$ space in Fig.~\ref{fig: ae} and \ref{fig: cenae}), for which we will assume 
 the observational bias should be negligible.
 For the $AP$,  we create a debiased
 population based on the work of Adams et al. (2014) and assume a number
 of bodies appropriate to 3.8 Gyrs ago using the work of Brasser and Morbidelli (2013).
In both cases we take the SSS model.

\begin{itemize}
\item $PP$: We study the orbital evolution of the present
 known Centaurs and TNOs for 50 Myrs. We consider the known objects whose
orbits are in an interval of semi-major axis 5.5 au $< a \leq$ 80 au, 
and perihelion $q \leq 40 au$.
 Tiscareno and Malhotra (2003) consider $q=33$ au
 a good boundary for TNOs having a similar dynamical lifetime
 to the Centaurs, but in order to be sure we go further in semi-major axis range till 40 au.
 The $PP$ sample is shown in $a-e$ and $a-i$ space in Fig. ~\ref{fig: ae}.
The orbital elements of these bodies were taken from the JPL-SBDSE. 
We then create  10 clones for each Centaur, and 5 for each TNO
 with these osculating elements:
 $a=a_0\pm0.005$, $e=e_0\pm0.003$ and $i=i_0\pm0.01$,
 similarly to Horner et al. (2004a), where
 $a_0,e_0$ and $i_0$ are the initial elements.
These clones are evolved over 50 Myrs in the SSS
  during which we check which bodies reach the
 main belt region, and how many asteroids and which main belt
 families are perturbed.

\item $AP$: Schwamb et al. (2014) estimates that the total number of
  the large bodies ($m>10^{-9} m_\odot$, which we take to mean those bodies 
with  visual r-magnitude less than 19.5, e.g. Gladman et al. 2001), we are
  interested in is 12 bodies of which 9 are currently
  known. Extrapolating from the results from Brasser and Morbidelli (2013), we
  anticipate that 3.8 Gyrs ago, the TNO population was about 17 times
  more numerous than now.  So we take the size of the ancient
  population to be $12*17=204$ TNOs here. Eris' mass is picked as
  representative of the largest sizes expected in the primordial
  Kuiper Belt, $m_{Eris}=1.66\cdot 10^{22} kg \sim8\cdot 10^{-9} m_\odot$ (Brown and Schaller 2007).

  We simulate this ancient population assuming the debiased
  orbital distribution of Adams et al. (2014). This assumes that the
  current distributions in sub-groups of the TNOs are in the same
  proportions as the post-LHB ones. From the Deep Ecliptic Survey ($DES$), Adams et al. (2014)
  computed that only 3 classes of TNOs can contain objects
  as massive as we desire: the Classical, the Scattered and the $3:2$ resonant.
  These relative populations of large objects (at least at $H<4$) can
  be approximated by a 2:2:1 ratio among them, which we assume
  represents approximately the post-LHB situation.  We choose a number
  of C+TNO clones at the previous ratios in sufficient numbers to have
  a reasonable statistical sample among the principal osculating
  elements ($a$, $e$ and $i$), following Adams et al. (2014). Our initial
  population was 170 clones for the Classical and Scattered population
  and 85 for the $3:2$ population.

 We integrate the orbits of this synthetic population for 200 Myrs.
\end{itemize}


\section{Interactions with the main belt asteroids: numerical estimates of the diffusion into the inner Solar System}

\label{orbevo}
We now describe the numerical results considering the present and the ancient
 populations:
\begin{description}
\item[$PP$:]
C+TNOs start to cross Jupiter's orbit and enter in the main belt 
region after experiencing close encounters
 (exciting their eccentricity) with the 
giant planets: 
 Jupiter is the most important for the 
Centaurs, and Neptune for the TNOs. 
 It is interesting to note that initially 
the largest Centaurs' orbits are beyond 
Saturn and in particular beyond the  J4:1 resonance (Fig.~\ref{fig: ae}).

 We find in the numerical integrations that solely those with $q<34$ au can 
enter in the main belt region in 50 Myrs of evolution, with the exception of C+TNOs in resonance like N2:3, which can have larger perihelia. 

In Fig.~\ref{MBAs} we can see 2 examples of 100-km size bodies that
enter the belt, passing through it for some kyrs and eventually
escaping: a Centaur (the first one which escapes) and a TNO (a Kuiper
Belt Object). The third (and last) object, in Fig.~\ref{MBAs},
 which crosses Jupiter's orbit is initially a member
 of the Scattered Disk which arrives only next to Jupiter's orbit
 after an important close
encounter with Saturn and is ejected out of the Solar System.

The majority of the Centaurs that enter in the Main Belt pass through
 in the first $\sim 10$ Myr (average value) and stay there for
 about $\sim$120 kyr. 

 They usually end as Sun-grazers or collide with the Sun or escape.
  Some stay in the belt for as long as $\sim 3$ Myr.

  TNOs reach the asteroid belt later (as expected, since they are Centaur
progenitors), at $\sim17$ Myr (average time)
 and reside there for a period of time similar to the Centaurs' one. 
 In general, C+TNOs can arrive through all the 50 Myrs of integration.
240 Centaur clones out of 1023 enter the main belt,
and 73 TNOs out of 2871,  respectively 23\% and $\sim3$\%.
 Table~\ref{totlife} shows the time of residence in the main belt.

 The C+TNOs that stay there the least time
 usually experience several close encounters with Jupiter,
 which modify their orbits substantially (see previous Section).
 The Centaurs which enter in the main belt are those ones that start their
  journey outside Saturn: their initial
 semi-major axis is $a\gtrsim13.7$ au. Some Centaurs are found to have a very
 high probability to interact with the belt, i.e. 
2003 QC$_{112}$ and also the largest one, 1995 SN$_{55}$.\\

 About 100 MBAs out of 1582 have close encounters with TNOs ($\sim
 6$\%) and only 5 ($<1$\%) of these produce a drift of the order of
 10$^{-4}$ au.  There are also a few cases where asteroids have
 multiple deflections of the order of $10^{-5}$ au, by the same C+TNO
 (even though their mass is only of the order of $10^{-10} m_\odot$).

\begin{table}[!h]
\begin{center}
\caption{C+TNOs total life (Myrs) in the main belt (maximum, $T_{max}$ and
 average value,  $<T>$) and (average) arrival time ($T_a$).
The symbol * underlines that the synthetic populations of TNOs ($AP$) was
 integrated for longer time than the present population ($PP$),
 200 Myrs instead of 50  Myrs.}
\begin{tabular}[h]{|l|c|c|c|}
\hline
\hline
Group    &    $T_{max}$    & $<T>$    & $T_a$  \\
\hline
\hline
Centaurs ($PP$)  &  3.1 & 0.1  & 10.0  \\
\hline
TNOs ($PP$)  & 1.9 & 0.1 & 16.8\\
\hline
TNOs ($AP$)* & 3.7 & 0.2 & 85.2 \\
\hline
\hline
\end{tabular}
\label{totlife}
\end{center}
\end{table}

\begin{figure}[!h]

  \centering
  \centering \includegraphics [width=0.45\textwidth]{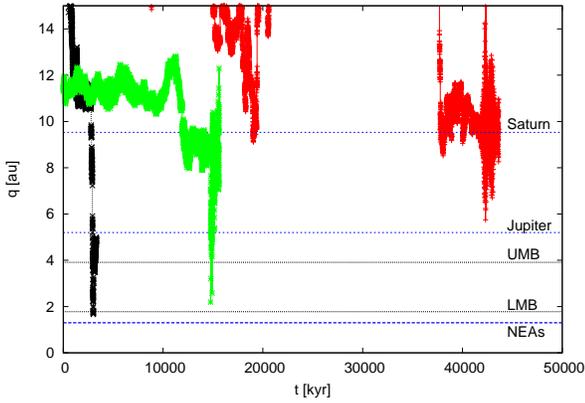}

\caption{Perihelia of a sample of 100 km-size bodies which cross Jupiter' orbit,
 during their 50 Myr  evolution and eventually enter in the main belt.
 UMB $=$ Upper border of the Main Belt, LMB $=$ Lower 
border of the Main Belt, NEAs $=$ Near-Earth Objects.}
\label{MBAs}
\end{figure}

\item[$AP$:]

From our numerical integrations, we find that in 200 Myrs
only objects from the Classical and 3:2 populations, 
 can enter in the main belt. They do so after entering in the 
 Centaur region and interacting with the giant planets. 

The synthetic TNOs in this sample which reach the main belt stay
 about 1/4 million years. 
8.5\% (36 out of 225)  of the TNOs simulated
enter the main belt in 200 Myrs, and a good
 number of them ($\sim$50\% of the ones which enter)
 provide an effective perturbation on the MBAs, see the histograms with the
lowest rates in Fig.~\ref{fig: decay}.
This last figure shows that the main belt entry rate of TNOs (the $TNOs_M$)
 and the sub-group of them  which not only enters but also makes an
 effective perturbation on MBAs ($TNOs_{M'}$) is similar, meaning that
 the large bodies which make a scattering and the ones which do not, 
have a similar decay in population.
 
Also shown is the rate at which TNOs (the 'survivors', $TNOs_S$:
all the TNOs which did not have ejections or impacts).
  Fitting the trends,
 with an exponential decay function,
 $N_{TNO}=N_0e^{-t/\lambda}$, we can compute
 their half-life ($t_{1/2}=\lambda ln 2$):
  $t_{1/2,TNOs_M}=215\pm132$ kyr and $t_{1/2,TNOs_{M'}}=125\pm28$ kyrs.
 The decay rate of $TNOs_S$ is at least 3 times slower,  
$t_{1/2,TNOs_{S}}=847\pm69$ kyrs, however, all rates are of the same order.

Concerning the most perturbed main belt asteroids, that is,
 those with $\Delta a \gas K_{kill}$, 
we find 66 deflections this large out of the 261 recorded, $\sim$25\%.
The largest reaches $\sim 10$ times $K_{kill}$.
Of these deflections, 7 were Vestoids ($V$, $N_V=7$) and 59 non-Vestoids
 ($NV$, $N_{NV}=59$).
 In order to compute the real number of asteroids with $H<14$ deflected 
in 200 Myrs, we have to scale these numbers by the real number of TNOs
 3.8 Gyrs ago. In the case of the $NV$, also by the real total
 number of the main belt $NV$ with $H<14$, $N_{NV,T}=26183$: 
(the total number of non-Vestoids with $H < 14$,  see Section~$\ref{model}$).

The total number of deflected asteroids will be the sum of the total $V$ ($N_{T,V,def}$) and $NV$ ($N_{T,NV,def}$) deflected: 

\begin{eqnarray} \label{eq:10}
N_{T,def} &=& N_{T,V,def}+N_{T,NV,def} \\
&=& N_V S_1 + N_{NV} S_2 S_1 \\
&=&\frac{N_V N_{T,1}}{N_{deb}}+N_{NV}\frac{N_{NV,T}}{N_{NV,S}}\frac{N_{T,1}}{N_{deb}}=707
\end{eqnarray}
 where $N_{NV,S}$ is the total number of $NV$ in the sample used for the numerical
 integrations. Then $N_{T,1}=204$  ($12\times17$)
is the total number of the most massive ($m>10^{-9} m_\odot$)
TNOs at 3.8 Gyrs ago and $N_{deb}$ is the number of synthetic TNOs considered
 for the orbital  evolution of the debiased population
 ($N_{deb}=170+170+85=425$).  $S_1$ and  $S_2$ (where $1=TNOs$ and $2=MBAs$)
 are the appropriate scaling factors for quantities respectively dependent
 on the real number of TNOs and of MBAs.
Having 707 km-size asteroids deflected out of their families
 means that about 3\% of the MBAs have their orbits significantly changed
 by the passage of the C+TNOs in 200 Myrs, and in particular the percentage, therefore the perturbation, will be higher in specific regions of the main belt (so also in some specific family), see Section \ref{famper} and
 in particular Fig.~\ref{fig: driftfamilies}.

\begin{figure}[!h]

  \centering
  \centering \includegraphics [width=0.45\textwidth]{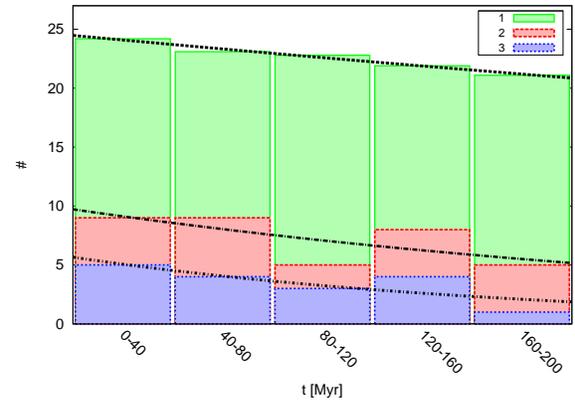}
\caption{Number of TNOs (\#) per interval of time (t): rate every 40 Myrs.
The histogram represents 3 populations rates:
 (1) surviving TNOs (columns with the highest rates. This rate was scaled by 10 in order to show better the trend of the other 2 groups)  for the classes which contribute to the scattering in the main belt: Classical and N2:3 classes; (2) TNOs entering the main belt and  (3) TNOs entering the main belt 
which provide a $\Delta a> K_{kill}$. Each group is fitted to its own decay function.}
\label{fig: decay}
\end{figure}

\end{description}

\subsection{Typical dynamical orbit of the belt-crossing Centaurs and TNOS}
\label{dynamo}


\begin{figure}[!h]
  \centering
  \centering \includegraphics [width=0.45\textwidth]{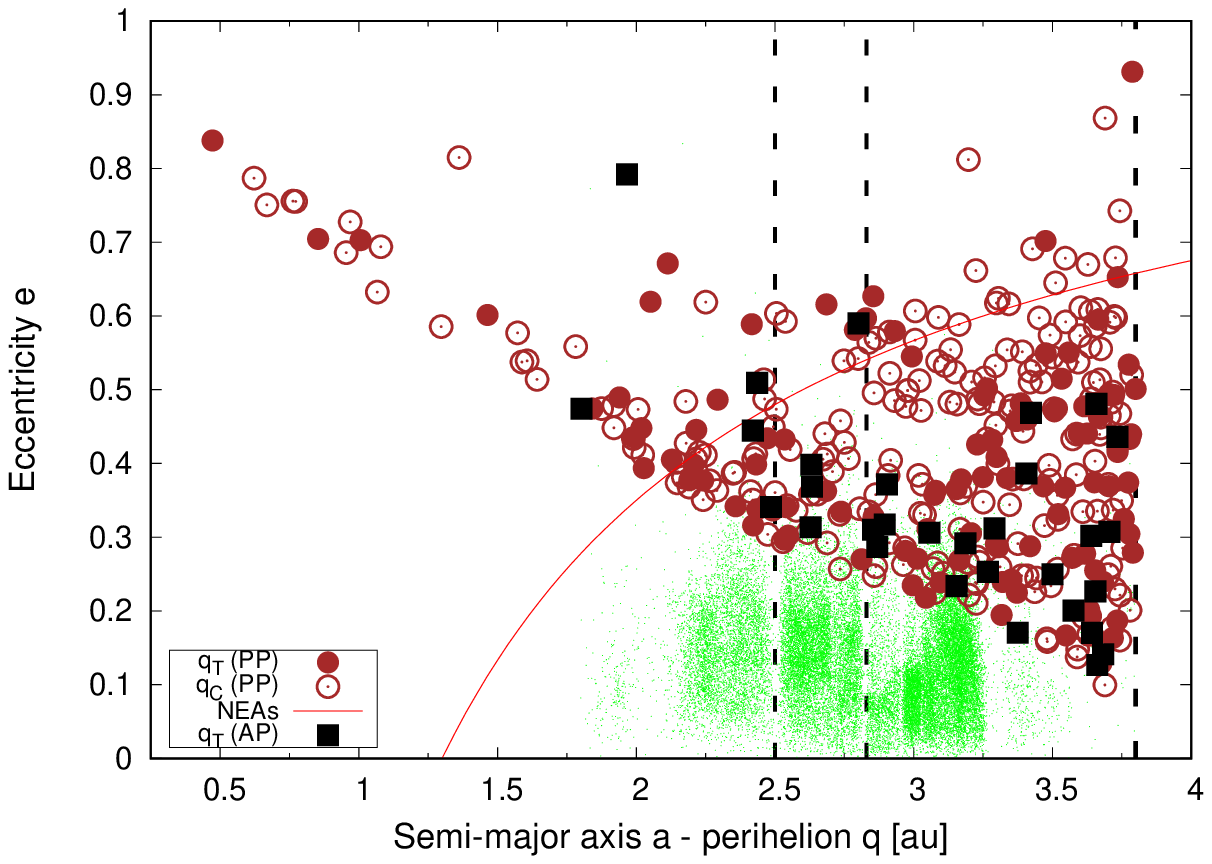}
\includegraphics [width=0.45\textwidth]{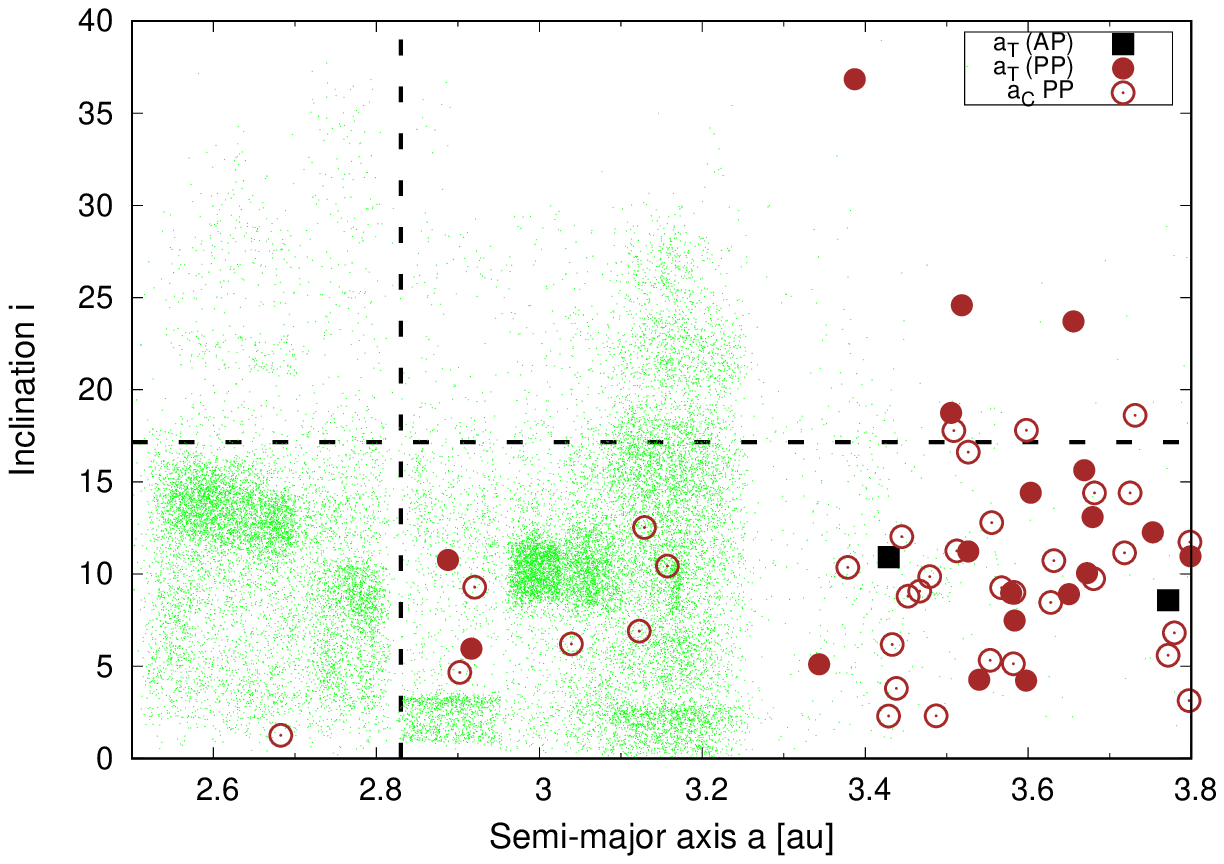}
\includegraphics [width=0.45\textwidth]{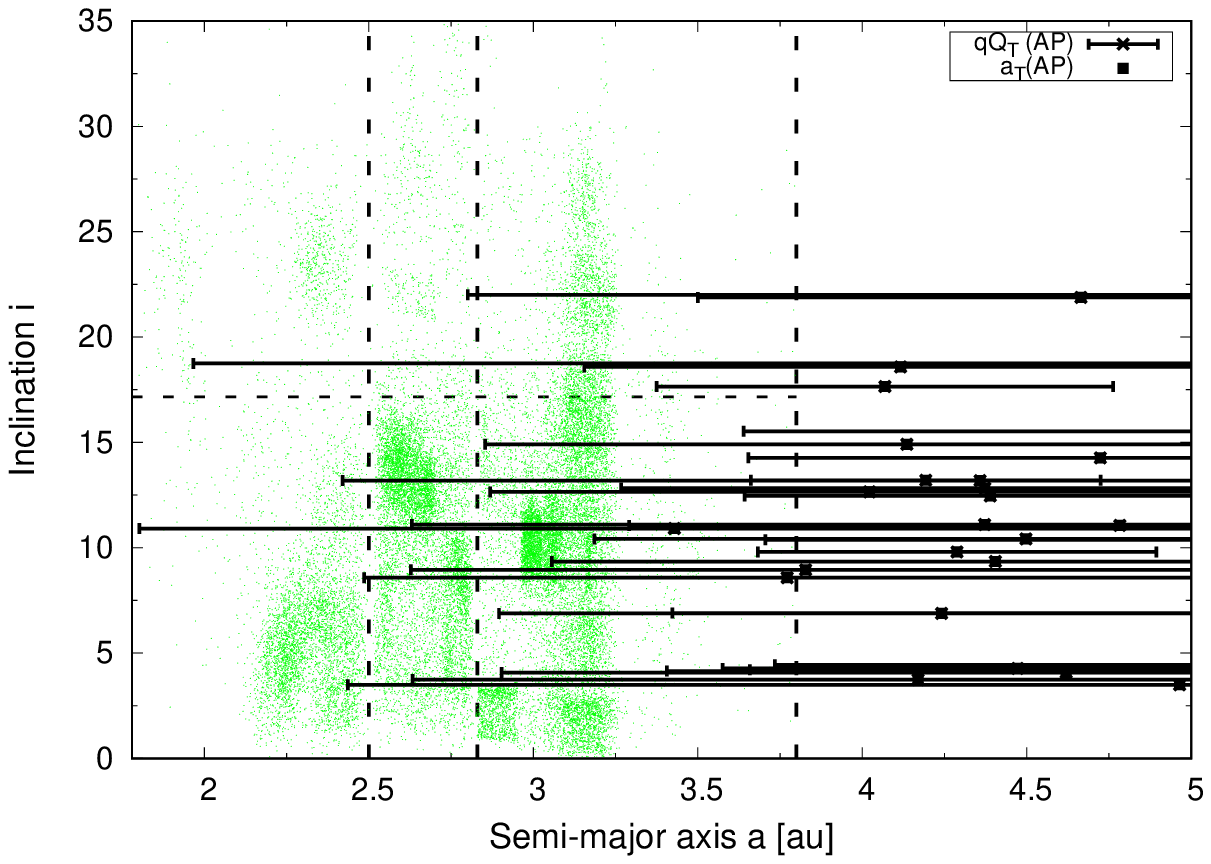}
\caption{In each panel, the semi-major axis of the 
main belt asteroids with $H<14$  is represented  with small green
 dots. The larger points are related  to the C+TNO populations
 when they first enter the main-belt.
Top panel: overplotted is the perihelion versus eccentricity projection
 of the Present Population (Centaurs, $q_C$ (PP), and TNOs, $q_T$ (PP)) and 
of the Ancient Population ($q_T$ (AP)). 
 Vertical dotted lines are the borders of the main belt regions. The region to the left of the red line is the NEA region.
Middle panel: a semi-major axis  versus inclination projection of the Ancient 
 and Present populations.
Bottom panel: semi-major axis ($a_T$ with aphelion and perihelion limits, $qQ_T$) versus inclination projection of the TNOs of the Ancient Population.}
\label{fig: dynbelt}
\end{figure}

The C+TNOs initially interact more with the outer main belt than 
  the inner regions of the belt. Fig.~\ref{fig: dynbelt} shows this behaviour
 clearly. They have semi-major axes which spread from the main belt to outer regions larger than 5 au.
  Only a small number of them has $i>17.16^\circ$; however, as shown in Fig.~\ref{fig: driftfamilies}, some families at high inclinations can be influenced by the TNOs.
The C+TNOs arrive in the main belt at high eccentricities
 (in comparison with typical MBAs),  generally $e\gtrsim 0.16$,
 and inclinations, $i\las25^\circ$, with only one case at $i=36^\circ$.
In particular the typical semi-major axis of the Ancient Population,
 when arriving at the belt, favors the outer belt beyond 3.4 au,
 while on the contrary the Present Population can arrive
 also in the middle main belt.
For these large bodies entering in the belt, the average and median values of their 
semi-major axes are respectively  $(5.7\pm2.6)$ au and  5.0 au in the $PP$ of  Centaurs,
 and 
 $(5.7\pm4.8)$ au and 4.9 au of TNOs, respectiverly.
However, considering the whole Centaurs and TNOs, these values become 
 $(5.7\pm3.6)$ au and 4.9 au in the $PP$ and $(4.9\pm1.2)$ au and 4.4 au in the $AP$.

The ancient population is the most important one in term of scattering
 MBAs out of their relative main belt families. The most perturbed regions of 
 the belt are shown in the lower panel of  Fig.~\ref{fig: dynbelt},
 which illustrates the perturbed region of the $AP$ by a line connecting aphelion and perihelion. The most influenced region is the Outer main belt at low inclinations.

Next section will show in more
detail which asteroid families are most affected. Here, in particular
 we want to underline that the $AP$ can reach inclinations up to
$23^\circ$ and have a perihelion in the inner main belt.
 
The AP orbits are usually controlled by Jupiter which also affects their eccentricities through close encounters, and which eventually ejects the C+TNOS in hyperbolic orbits or causes them to collide with the Sun. 
 Although the behaviour between the AP and PP is quite similar,
 the main difference is that on average the Present Population has
 lower perihelia when it starts to enter in the belt and, 
in fact , the median eccentricity of the $PP$,
  $\bar{e}_{AP}=0.39$, $\bar{e}_{AP}=0.31$, is higher
 than the one of the $AP$, $\bar{e}_{AP}=0.31$.
 This also means that the $AP$ has more possibility to encounter
 Main belt asteroids, because the related orbits intersect those of the MBAs
 for  longer time.

Finally, we note that some\footnote{For the Present Population, 16\% of the Centaurs and 18\% of the TNOs, for the Ancient Population 6\%. The rest of all those populations which interact with the main belt (having perihelion, $q\las3.8$ for at least 2 kyrs, see Section. 2)  has semi-major axis $a>3.8$ au. Centaurs
 and Trans-Neptunian Objects reach orbits similar orbits
 to those of the present main belt asteroids (see in particular the
 middle panel of Fig. \ref{fig: dynbelt})
 and thus there could be Centaurs and/or TNOs interlopers
 in the main belt
 (as suggested e.g. by Fern\'andez and Sosa 2015). Similar orbits to our
 simulated C+TNOs interacting with the belt are in particular
 the orbits such as asteroids
 (511) Davida, (52) Europa, (31)  Euphrosyne, and (24) Themis in the outer belt.}

. 

\subsection{Can C+TNOs diffuse asteroid families?}
\label{famper}

 The effect of asteroidal diffusion by C+TNOs
 is very small for the case of young families
(the $PP$ results):
 only one case of a deflection over $K_{kill}$ by a TNO was recorded.


Therefore, perturbations by the C+TNOs
 have a negligible effect on asteroid families for the current main belt.
In the $AP$ case, there is more time (3.8 Gyr)
 for C+TNOs to be effective at disrupting asteroid families,
 and so older families might still be affected.
  Numerical integrations show that the time spent
 by individual C+TNOs within the main-belt 
is different from past to present, at least the double of
 the average for the $AP$ versus the $PP$. 
This can be seen in the orbital evolution of our TNO population
 (Fig.~\ref{fig: decay}) and was shown also by Brasser and Morbidelli (2013).
  Most of the $AP$ bodies enter the main belt
 sooner with lower eccentricity than $PP$ (see Subsection \ref{dynamo}),
 so they can spend more time in the belt
 and with a smaller encounter velocity, which helps
 to increase the deflection of the asteroids. 
The TNO population decay  also explains why we do not see important
 perturbations of the main belt now:
 the TNOs that pass now through the solar system are on more eccentric orbits
 than in the past.

\begin{figure*}[!!h]

  \centering
  \centering \includegraphics [width=0.9\textwidth]{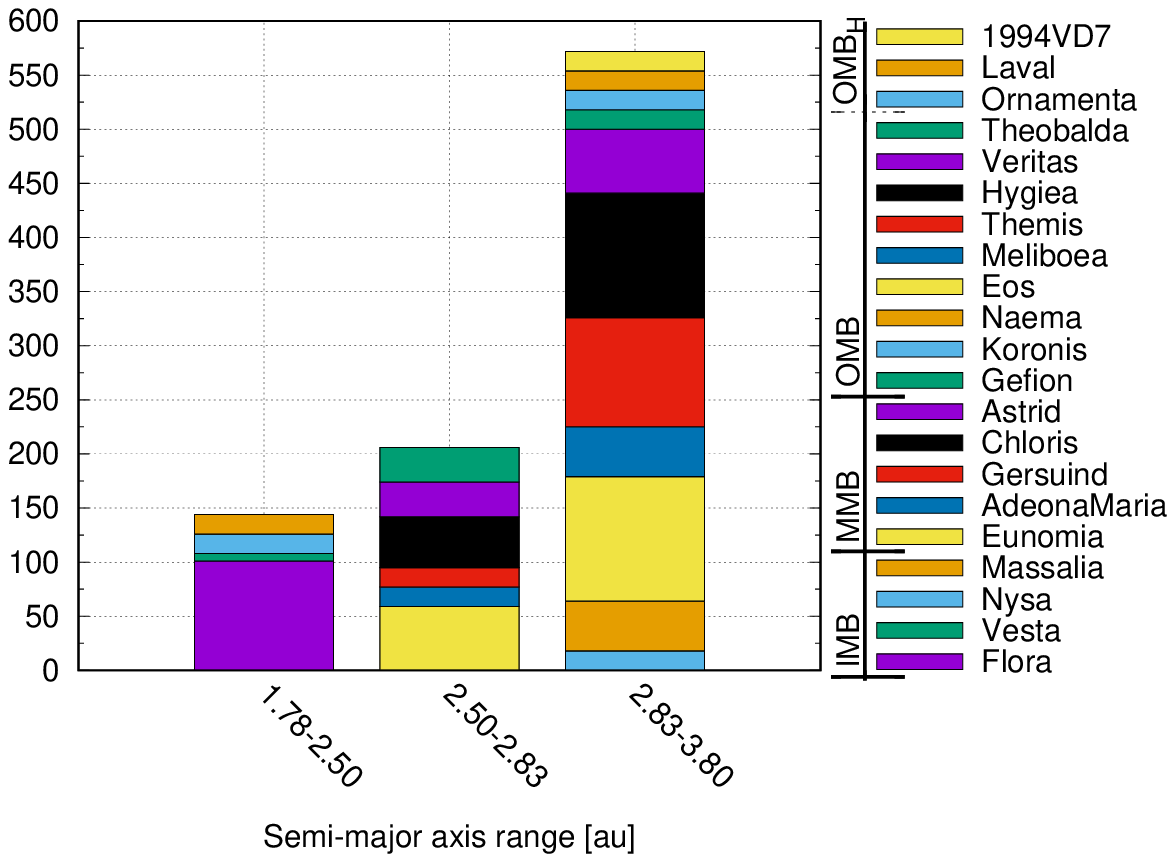}
\caption{Number of asteroids that drifted more than
 $\Delta \geq 3\cdot 10^{-4}$ au by family,
 divided into the 3 regions of the asteroid belt
 (by semi-major axis):
 1.78 au $< a <$ 2.50 au (IMB, defined before in Table 1 and so on for the other next 3 groups): Flora, Vesta, Nysa,
Messalia; 2.50 au $\leq a <$ 2.83 au (MMB): Eunomia, Adeona-Maria, 
Gersuing, Chloris, Astrid, Gefion; 2.83 au $\leq a \leq 3.8$ au (OMB):
 Koronis and then Naema, Eos, Meliboea, Themis, Hygiea, Veritas,
 Theobalda, Ornamenta, Laval and 1994 VD$_7$ (only this last 3  
are highly (H)
inclined families: $i>17.16^\circ$).}          
\label{fig: driftfamilies}
\end{figure*}

Because the fraction of MBAs affected over 3.8 Gyr at the current rate would be
at least 3\%, this effect can not be neglected in the long-term.
C+TNOs reached the main-belt in much
increased numbers in the past and so they have participated in the
disruption and diffusion of very old families.
Different main belt families are expected to have experienced the deflection of at least 
100 family members with sizes of the order of one km beyond the cutoff
 $\Delta a\gas K_{kill}$ due to the the influence
 of the passage of the ancient population, and these are:
Hygiea (the most perturbed one if we consider also
 asteroids affected at the border of its region),
Eos and Themis, in the $OMB$, and Flora in the $IMB$.
On a smaller scale (more than 20 but less than 100 deflected asteroids
 of kilometer sizes, $H<14$):
Eunomia, Chloris, Astrid and Gefion in the $MMB$, and Eos, Meliboea and Veritas
 in the $OMB$. Fig.~\ref{fig: driftfamilies}, clearly show
 that the $OMB$ is the most affected region of the main belt.

 Although there is still some debate in estimating properly the ages of the
 asteroid families, much work has been done to establish their ages via
 dynamical properties of these populations,
 e.g. (Broz et al. (2013), Spoto et al. (2015) and Carruba et al. (2016). 
 These works assert that the eldest families, ones that might have existed
 in the first 200 Myrs after the end of the LHB are few, 
 i.e. Themis, Maria, Koronis, Ursula, Eunomia and Flora
 (only the last two families for Carruba et al. 2016).
Our results show that all these old families can be perturbed significantly,
 except for Ursula and Maria, which have a significant lower
 number of deflected asteroids with a $\Delta a > K_{kill}$. 
However, we point out that the computation of the age via dynamical
properties typically ignores the effects of C+TNOs close encounters. If
C+TNO effects are significant at early times, the currently-accepted
asteroid family ages of old families can be in part inaccurate,
 especially for the most affected zones of the Main belt by
 close encounters with large TNOs in the time span of about $\sim$3-3.8 Gyrs ago:
 the main result here is that in the first hundred million years after the
 $LHB$ main belt families can be significantly perturbed by massive Centaurs and TNOs.

\section{Conclusions}
\label{sec: concl}
Centaurs and TNOs do not cause significant
deflections with the main belt asteroids at the current time 
 but they can perturb them significantly in the time span
 from the LHB to the present, especially before 3.5 Gyrs ago.
Our results show that only C+TNOs with a mass larger than $10^{-9} m_\odot$
can produce an important change of a MBA-orbit, meaning a change
 $\Delta a \gas K_{kill}$ au.

Current C+TNOs that enter the main belt with a low semi-major axis have 
a very short average life inside the belt and they escape the soonest.
We find that the TNOs of the ancient population reside
 in the belt longer than the present population, almost twice the time.
 23\% of the Centaurs and 3\% of the TNOs 
enter in the main belt for at least 2 kyrs over the recent 50 Myr time span.
Some C+TNOs stay for relatively long periods in the main belt in our
 simulations, up to 3 Myrs (3.7 Myrs for the $AP$ case), with some
 low-eccentricity orbits, $e\sim0.1$.
Concerning Centaurs, we found that only ones with an initial semi-major axis larger than
 that of Saturn enter the main belt, and in particular, with $a\gtrsim13.7$ au.

 The orbits of TNOs during their belt-crossing phases resemble those
 of known large main-belt asteroids,
so this could be a suggestion to investigate in more detail whether some dark,
 primitive main belt asteroids on short-lived orbits may be former
 TNOs, e.g. some asteroids like (511) Davida show this behaviour.

Typical eccentricities and inclinations of the C+TNOs when they reside
 in the belt are respectively $e\gtrsim 0.16$ and $i\lesssim25^\circ$. 
 Most of the C+TNOs pass initially in the outer main belt (in particular the
 ancient population) and their 
semi-major axis is typically $\sim5$ au, with the present population having larger values on average, but more present interiorly than the ancient population.\\
Presently C+TNOs orbits pass rarely into the inner main belt,
 they pass mostly through the outer one, generally, with $a=2.6-3.8$ au.
It is important to note that, under our results, billions of years ago, after 
the Late Heavy Bombardmetnt,
 massive ($m>10^{-9} m_\odot$) TNOs would have entered into the
 inner solar system and with lower eccentricity on average than the present
 population, having more probabilities to interact with main belt asteroids.

 The young asteroid families currently can not be perturbed
 much by even the largest C+TNOs; the probability that massive
 bodies arrive now is also low or negligible.

On the contrary, at least 3\% of the main belt can be significantly perturbed
  in the first hundreds of million of years after the LHB.
 For this reason some old families and/or former ``paleo-families''
 which may have existed in the past can be modified after the $LHB$ by large minor bodies. 
 Under our model, Flora and Eunomia are the most affected families, therefore
their past populations were much larger.
If we suppose that the estimated age of other old families
 are biased by neglecting the effects of
 close encounters with the C+TNOs,
  we could assert that those families with underestimated
 ages might have  been perturbed also. Perhaps the Hygiea family,
 or a paleo-family of Hygiea:
 from our results a large part of the border of its region
 in osculating elements  was depleted by the passage of large C+TNOs. 
 The most affected region of the belt after the LHB was the
Outer main belt and its related families.
We lastly underline that Centaurs and TNOs can also disperse part of
 the main belt families into other dynamical
 regions, in an interval of time $t\gas100-200 Myrs$ after 3.8 Gyrs ago. They can contribute on diffusing out some asteroids with 
a typical taxonomic type of its own relative family and mix them with other
 asteroids of different taxonomic type inside other families,
 like the case of the C-types in the Hungaria family, 
and may give rise to small interlopers in families.

\acknowledgments
 MAG wants to acknowledge the support by the Austrian FWF project
P23810-N16 and  the ``Reitoria de p\'os-gradua\c{c}\~{a}o da UNESP'' (PROPg, grant PVExt-2015). 
The core part of this work was done  when MAG has been present at UNESP
 as "visiting fellow". MAG wants also to thanks Prof. V. Carruba for his important
suggestions for the paper and Dr. Y. Cavecchi for suggestions in
computational improvements and Prof. A. Morbidelli for data on the population decay of the TNOs. 
SA wants to thank Brazilian National Research Council (CNPq, grant 13/15357-1).
 This work was also supported in part by the
Natural Sciences and Engineering Research Council of Canada.
\label{lastpage}


\smallskip

\noindent

\nocite{*}
\bibliographystyle{spr-mp-nameyear-cnd}
\bibliography{biblio-u1}

\end{document}